\documentclass[twocolumn,amsmath,amssymb,prl,nofootinbib]{revtex4}
\usepackage{amsmath,epsfig}
\usepackage{amssymb,amsfonts}
\usepackage{latexsym}
\usepackage[latin1]{inputenc}
\usepackage{tocvsec2}
\usepackage{subeqnarray}
\usepackage{xcolor}
\usepackage{graphicx}
\usepackage{longtable}
\def\be{\begin{equation}}
\def\ee{\end{equation}}
\def\bea{\begin{eqnarray}}
\def\eea{\end{eqnarray}}

\newcommand{\nn}{\nonumber}

\begin{document}

\title{Holographic Pair and Charge Density Waves}

\author{Sera Cremonini}
\email{cremonini@lehigh.edu}
\author{Li Li}
\email{lil416@lehigh.edu}
\affiliation{Department of Physics, Lehigh University, \\ Bethlehem, PA, 18018, USA.}
\author{Jie Ren}
\email{jie.ren@mail.huji.ac.il}
\affiliation{Racah Institute of Physics, \\ The Hebrew University of Jerusalem, 91904, Israel.}

\begin{abstract}
We examine a holographic model in which a $U(1)$ symmetry and translational invariance are broken spontaneously at the same time.
Our construction provides an example of a system with pair-density wave order, in which the superconducting order parameter 
is spatially modulated but has a zero average.
In addition, the charge density oscillates at twice the frequency of the scalar condensate.
Depending on the choice of parameters, the model also admits a state with co-existing superconducting and charge density wave orders, in which the scalar condensate has a uniform component.
\end{abstract}


\maketitle

\section{Introduction and Discussion}

Over recent years holographic techniques originating from the AdS/CFT duality, and first developed in string theory,
have been used to analyze models that may be in the same universality class as many highly correlated systems.
Thanks to such approaches, challenging questions about dynamics in quantum phases of matter at strong coupling can be mapped to 
processes in theories of gravity that are tractable. Thus, holography provides a window into the often unconventional physics of these systems.

Inhomogeneities, striped phases and competing orders are believed to play an important role in the rich phase structure of high $T_c$ superconductors~\cite{Zaanen,Vojta:2009,EmeryKivelsonTranquada}.
In certain regions of the phase diagram -- such as the pseudo-gap regime -- many of these orders appear to be intertwined and sometimes have comparable strengths and common origin. Here we focus on a particular broken-symmetry phase, the pair-density wave (PDW)~\cite{Himeda:2002,Berg:2007}, 
in which charge density wave (CDW) and superconducting (SC) orders are intertwined in a very specific way, and in which spin density wave (SDW) order can also play a role. 
PDW phases seem to be a robust feature of models of strongly correlated electrons including high $T_c$ superconductors, and
 there is experimental evidence that they appear at least in the cuprate La$_{2-x}$Ba$_x$CuO$_4$~\cite{FradkinKivelsonTranquada,BergFradkinKivelson,Berg:2009dga}. 

In this paper we construct and study a holographic model which exhibits either  PDW or
co-existing SC+CDW orders, depending on the parameters in the theory.
To our knowledge this is the first holographic setup to realize a PDW.
While both PDW and SC+CDW break translational invariance and a $U(1)$ symmetry spontaneously,  
there is a key difference between them.
In a PDW the superconducting order parameter varies periodically as a function of position, but does so with a zero average,~\emph{e.g.} $\langle O_\chi\rangle \propto \cos(k\,x)$.
Moreover, in such a phase the charge density, which is also modulated, has a period which is~\emph{half} of that of the scalar condensate,
\emph{e.g.} $\rho(x) = \rho_0 + \rho_1 \cos(2k\,x)$. 
In contrary, a SC+CDW state has a uniform component to the condensate, which oscillates at the same frequency as the charge density.

In our construction the $U(1)$ and translational symmetries are broken spontaneously  at the same time.
The set-up we adopt includes, in addition to gravity, two real scalar fields $\chi$ and $\theta$ and 
two vector fields $A_\mu$ and $B_\mu$.
The couplings between the scalars and the gauge fields can be generated via the St\"{u}ckelberg mechanism.
Indeed, our theory is not of the form of the standard holographic superconductor~\cite{Gubser:2008px,Hartnoll:2008vx}, but rather
falls within the~\emph{generalized} class of models advocated for in~\cite{Franco:2009yz}.
The more general structure of the scalar couplings allows us to break the desired symmetries without the need to introduce additional fields. 

Here the presence of two vector fields (and the interaction between them) is crucial for obtaining the symmetry breaking features we are after. 
The role of the gauge field $A_\mu$ is transparent, since it provides a finite charge density $\rho_A$ whose modulations agree with the behavior of a 
PDW or CDW state.
What distinguishes whether the 
system is described by a PDW or by SC+CDW is whether the scalar $\chi$ is charged or not under the second vector field $B_\mu$.
The physical interpretation of $B_\mu$ depends on details of the model. In particular, when the field is massless it can be associated with spin degrees of freedom, 
and the modulations in its density $\rho_B$ could characterize SDW order. 

Before discussing our model we should mention that striped orders in holographic superconductors have been studied in a variety of setups, starting with~\cite{Flauger:2010tv}, 
in which an inhomogeneous phase was sourced by a modulated chemical potential. There have been many generalizations since then. 
In particular, a study of backreaction in the presence of a periodic potential was initiated in~\cite{Horowitz:2013jaa}. 
However, in these setups the breaking of translational invariance was explicit and not spontaneous. 
Holographic superconductors with spontaneously generated helical structure were reported in~\cite{Donos:2011ff,Donos:2012gg}.
The competition between superfluid and striped phases has been examined within the context of holography, see~\cite{Donos:2012yu,Cremonini:2014gia} for top-down models.
The spontaneous formation of striped order in a holographic model with a scalar coupled to two $U(1)$ gauge fields 
was first studied in~\cite{Donos:2011bh} and more recently in~\cite{Donos:2013gda,Ling:2014saa,Donos:2016hsd,Kiritsis:2015hoa}
(note that these models preserve the $U(1)$ symmetry).

Here we have extended such constructions by simultaneously
breaking both symmetries spontaneously, and focusing on the differences between a scalar condensate with PDW vs. CDW order.
Moreover, we have recently seen in a number of holographic models of strongly correlated electrons
the advantage of using multiple vector fields, 
as they typically lead to richer physics,~\emph{e.g.}~\cite{Kiritsis:2015hoa,Donos:2012js,Cremonini:2016avj,Seo:2016vks}.
In particular, such a picture was used to construct phase diagrams that are similar to those of high $T_c$ superconductors as well as other strange metal materials in~\cite{Kiritsis:2015hoa}.
Our construction provides a further example of this idea.
Note that while in our analysis the mass of the vector $B_\mu$ does not affect any of the physics in a qualitative way, it is expected to play a role for applications to transport. 
It would be interesting to study the effects of disorder on the PDW state, as well as the consequences of stripe order on the conductive properties of the system and  on 
fermion spectral functions.
We leave these questions to future work.
A more detailed analysis for this model will appear in~\cite{LongerPaper}.

\section{Holographic Setup}
We choose our theory $S=\int d^{4}x \sqrt{-g} \, \mathcal{L}$ to describe gravity coupled to 
two real scalar fields $\chi$ and $\theta$, and two vector fields $A_\mu$ and $B_\mu$,
\begin{eqnarray}\label{actions}
&& \mathcal{L}= \mathcal{R}  + \frac{6}{L^2} -\frac{1}{2} (\partial \chi)^2 -\frac{Z_A}{4} F^2
-\frac{Z_B}{4} \tilde F^2
-\frac{Z_{AB}}{2}
F \, \tilde F \nonumber\\
&&-\mathcal{K}(\chi)(\partial_\mu\theta-q_A A_\mu-q_B B_\mu)^2-\frac{m_v^2}{2}B^2- \frac{m^2}{2}\chi^2 ,
\label{Smatter}
\end{eqnarray}
with  $F_{\mu\nu}=\partial_\mu A_\nu-\partial_\nu A_\mu$ and $\tilde F_{\mu\nu}=\partial_\mu B_\nu-\partial_\nu B_\mu$ denoting the field strengths of the two vectors, and 
$F \tilde F=F_{\mu\nu}\tilde F^{\mu\nu}$ for short.
We take the gauge field couplings $Z_A,Z_B,Z_{AB}$ to depend on $\chi$, and in particular
chose them so that in the limit $\chi\rightarrow 0$ they take the form
\begin{equation}
\begin{split}
\label{coupling}
&Z_A=1+\frac{a}{2}\chi^2+   \mathcal{O}(\chi^3) \, ,  \quad  Z_B=1+ \mathcal{O}(\chi^2)\, , \\
&Z_{AB}=c\, \chi + \mathcal{O}(\chi^2)\,  ,  \quad  \mathcal{K}=\frac{1}{2}\chi^2+\mathcal{O}(\chi^3)\, , 
\end{split}
\end{equation}
with $(a,c)$ constants. 
We note that the
$c$ parameter which controls the interaction $\sim Z_{AB}$ between the two fluxes will play 
a crucial role in the breaking of translational invariance.

While in general we will assume that $\chi$ is charged under both $U(1)$ fields, we will see that the $q_B=0$ case plays a special role, as it is associated with a PDW condensate.
On the other hand, $q_B \neq 0$ will describe a state with SC+CDW order.
Finally, note that while the current dual to $A_\mu$ is conserved, the same is not always true for the current dual to $B_\mu$, because of the 
mass term $m_v^2$. Although in this paper we consider both massless and massive cases for the sake of completeness, they lead to the same qualitative results. 
On the other hand the mass parameter $m_v^2$ is expected to affect \emph{e.g.} the transport properties of the system, which we plan to study in future work.

We are interested in considering two classes of background solutions to this system.
The first one is the electrically charged AdS Reissner-Nordstr\"{o}m (AdS-RN) black brane only supported by $A_\mu$,
\begin{equation}\label{RNads}
\begin{split}
&ds^2=\frac{1}{f(r)}dr^2- f(r)dt^2+\frac{r^2}{L^2}(dx^2+dy^2)\, ,\\
&f(r)=\frac{r^2}{L^2}\left(1-\frac{r_h^3}{r^3}\right)+\frac{\mu^2 r_h^2}{4r^2}\left(1-\frac{r}{r_h}\right)\, ,  \\
&A_t=\mu\left(1-\frac{r_h}{r}\right)\, ,
\end{split}
\end{equation}
where $r_h$ is the horizon, $\mu$ the chemical potential and other fields are trivial.
This background will describe the high temperature phase in which the dual theory possesses a global $U(1)$ symmetry, associated with the gauge field $A_\mu$.
The black brane temperature reads
$T=\frac{12 r_h^2-\mu^2 L^2}{16\pi L^2 r_h}$,
and in the extremal limit $T=0$ the near horizon geometry becomes that of $AdS_2\times R^2$, 
\begin{equation}\label{ads2}
\begin{split}
ds^2=\frac{L^2}{6 \tilde{r}^2 }d\tilde{r}^2-\frac{6 \tilde{r}^2}{L^2}dt^2+\frac{r_h^2}{L^2}\, d\vec x^2\, ,
\; \; A_t=\frac{2\sqrt{3}}{L}\tilde{r}\, ,
\end{split}
\end{equation}
with $\tilde{r}=r-r_h$ and the $AdS_2$ radius $L_{(2)}=L/ \sqrt{6}$.

We will then examine solutions with a non-trivial profile for $\chi$ and $B_\mu$.
These will describe the formation of a scalar condensate in the low temperature regime of the dual field theory, and provide holographic 
probes of phases with a broken $U(1)$ symmetry. 
Moreover, by allowing for modes which source spatial modulations, we will trigger instabilities to striped superconducting phases.
The detailed structure of the modulations of the condensate and charge densities will be sensitive to $q_B$ as well as the parameters in the theory, as we will see shortly.

\section{Striped Instabilities}
\label{section:spatial}

To determine whether in this model we can spontaneously break translational invariance at the same time as the $U(1)$ symmetry,
we need to examine the spatially modulated static mode in the spectrum of fluctuations around the unbroken phase.
Our strategy will be to first consider instabilities arising from the IR $AdS_2\times R^2$ geometry, and to construct analytically 
momentum-dependent modes which violate the IR $AdS_2$ BF bound.
The presence of such modes is
a strong indication that there should be a region in which one has superconducting 
order that is spatially modulated -- a striped superconductor. 
We will then move on to studying numerically the behavior of the perturbations and of the condensate at finite temperature.

\subsection{Instabilities of the IR $AdS_2\times R^2$ geometry}

We are now ready to examine  instabilities of the electrically charged AdS-RN black blane~\eqref{RNads}.
We start from the $AdS_2\times R^2$ background~\eqref{ads2} which arises as the IR limit of the zero temperature AdS-RN geometry,
and turn on the following two spatially modulated perturbations,
\begin{eqnarray}
\label{fluctuations}
\delta \chi =  \varepsilon\, w(r)\cos(k\,x),\quad \delta B_{t} =  \varepsilon\,  b_{t}(r)\cos(k\,x)\,,
\end{eqnarray}
where we have relabeled $\tilde{r} \rightarrow r$ for convenience, and 
$\varepsilon$ is a formal perturbative expansion parameter.
By substituting into the equations of motion and working at linear level in $\varepsilon$,
we obtain the two coupled equations
\begin{eqnarray}
\frac{6}{L^2} (r^2 w^\prime)^\prime-\frac{2\sqrt{3}\, c}{L}b_t'-  \left(M_{(2)}^2   +  \frac{k^2 L^2}{r_h^2} \right) w& = & 0 \, , \\
\frac{6 r^2}{L^2}b_t''+\frac{12\sqrt{3}\, c\, r^2}{L^3}w'-  \left(m_v^2+\frac{k^2 L^2}{r_h^2}\right) \, b_t &=&0 \, ,
\end{eqnarray}
with $ M_{(2)}^2 = m^2-\frac{6 a}{L^2}-2  q_A^2 $ and
$(a,c)$ as defined in equation~\eqref{coupling}.
We make the further ansatz
\begin{equation}
w(r)=v_1\, r^\lambda\,,\quad b_t(r)=v_2\, r^{\lambda+1}\,,
\end{equation}
where $v_1,v_2$ are constants and $\lambda$ denotes the scaling dimension of an IR operator in the one-dimensional CFT dual to the $AdS_2$ geometry.
The linearized  equations can then be written in matrix form, solving which we find
\begin{equation}
\lambda_+^\pm=-\frac{1}{2} + \sqrt{\frac{1}{4}+ m_\pm^2} \, ,
\;\;
\lambda_-^\pm= -\frac{1}{2}  -   \sqrt{\frac{1}{4}+ m_\pm^2} \, ,
\end{equation}
with
\bea
&& m_\pm^2 =\frac{L^2}{12}  \left[ M_{(2)}^2 +m_v^2 +12 \frac{c^2}{L^2}+24 k^2  \right]  \nn \\
&& \pm  \frac{L^2}{12}  \sqrt{(M_{(2)}^2 -m_v^2)^2+  24 \frac{c^2}{ L^2} (M_{(2)}^2 +m_v^2  +  6 \frac{c^2}{L^2}  +24 k^2)}  \nn
\eea
where we have fixed the chemical potential to be $\mu=1$.

The onset of the instability associated with the violation of the $AdS_2$ BF bound is linked to
$\lambda$ becoming imaginary, \emph{i.e.} when $ m_-^2<-\frac{1}{4} $.
For striped instabilities, one needs a non-zero wave number $k$ at which the value of $\lambda$ is imaginary, 
for a fixed choice of Lagrangian parameters. 
By inspecting the form of $m^2_-$, one can check explicitly that this is clearly possible for various parts of the parameter space.
As a specific example, for the parameters chosen in the finite temperature analysis below (\emph{e.g.} $ m^2 = -8, q_A =1, m_v^2 =0, L=1/2, a=4, |c|=2.34 $), we find a momentum range $ 0.99< |k| < 2.62$ in which the modes violate the BF bound, and are 
associated with spatially modulated phases. We come back to this point in greater detail in the numerical analysis below.

\subsection{Critical Temperature}

The instabilities of the IR $AdS_2$ solutions 
that we have just discussed 
occur at zero temperature.
Nevertheless,  they suggest that analogous instabilities should appear in the black brane background~\eqref{RNads} at finite temperature.
Next, we shall calculate the critical temperature $T_c$ below which the AdS-RN geometry becomes unstable, as a function of wave number $k$. 
In particular, if the scalar field instabilities are associated with a finite value of $k$, we will have found a striped condensate.
Note that to obtain $T_c$ it is sufficient to work to linear order in perturbations.

Motivated by the $AdS_2$ analysis, we turn on the same fluctuations as in~\eqref{fluctuations}.
By expanding around the AdS-RN background, one then obtains two coupled linear ODEs,
\begin{eqnarray}
&& w''+\left(\frac{2}{r}+\frac{f'}{f}\right)w'+\frac{c \mu r_h}{r^2 f}b_t'  \nn \\
&&-\frac{1}{f}\left(m^2-\frac{\kappa\, q^2\mu^2 (r-r_h)^2}{r^2 f}+\frac{k^2 L^2}{r^2}-\frac{a \mu^2 r_h^2}{2r^4}\right)w = 0\,, \nn \\
&& b_t''+\frac{2}{r} b_t'+\frac{c \mu r_h}{r^2}w'-\frac{1}{f}\left(m_v^2+\frac{k^2 L^2}{r^2}\right)b_t = 0 \, ,
\end{eqnarray}
which can be solved numerically.
We demand the fluctuations to be regular at the horizon at $r=r_h$, with
\begin{equation}
w(r) = w^h+\mathcal{O}(r-r_h)\,, \; b_t(r)=b_t^h(r-r_h)+\mathcal{O}(r-r_h)^2 \, . \nn
\end{equation}
On the other hand their $r\rightarrow \infty$ UV expansion is
\begin{equation}
\begin{split}
&w(r)=\frac{w_s}{r^{3-\Delta_\chi}} (1+ \cdots) + \frac{w_v}{r^{\Delta_\chi}}(1+\cdots)\,, \\
&b_{t}(r) =\frac{b_s}{r^{2-\Delta_B}}(1+\cdots) +\frac{b_v}{r^{\Delta_B-1}}(1+\cdots)\,,\;
\end{split}
\end{equation}
where the quantities $\Delta_\chi=\frac{1}{2}(3+\sqrt{9+4m^2L^2})$ and $\Delta_B = \frac{1}{2}(3+\sqrt{1+4m_v^2L^2})$ are, respectively, the scaling dimensions of the scalar operator dual to $\chi$ and vector operator dual to $B_\mu$.
Since we are only interested in breaking both symmetries spontaneously, we turn off the parameters $w_s$ and $b_s$, which correspond to the sources for the operators
in the dual field theory.

After fixing theory parameters, for a given wave number $k$ we expect there to be a normalizable zero mode appearing at a particular temperature.
We choose the $\chi$ mass term to be $m^2 L^2 =-2$, so that $\Delta_\chi = 2$, and consider two separate cases for the second vector field $B_\mu$.
We first take it to be massless,  $m_v^2=0$, so that  $\Delta_B =2$ and the associated current is conserved.
We then consider the case in which it is massive, choosing $m_v^2L^2=0.11$ in our numerics, corresponding to $\Delta_B =2.1$.
\begin{figure}[ht!]
\begin{center}
\includegraphics[height=.153\textwidth]{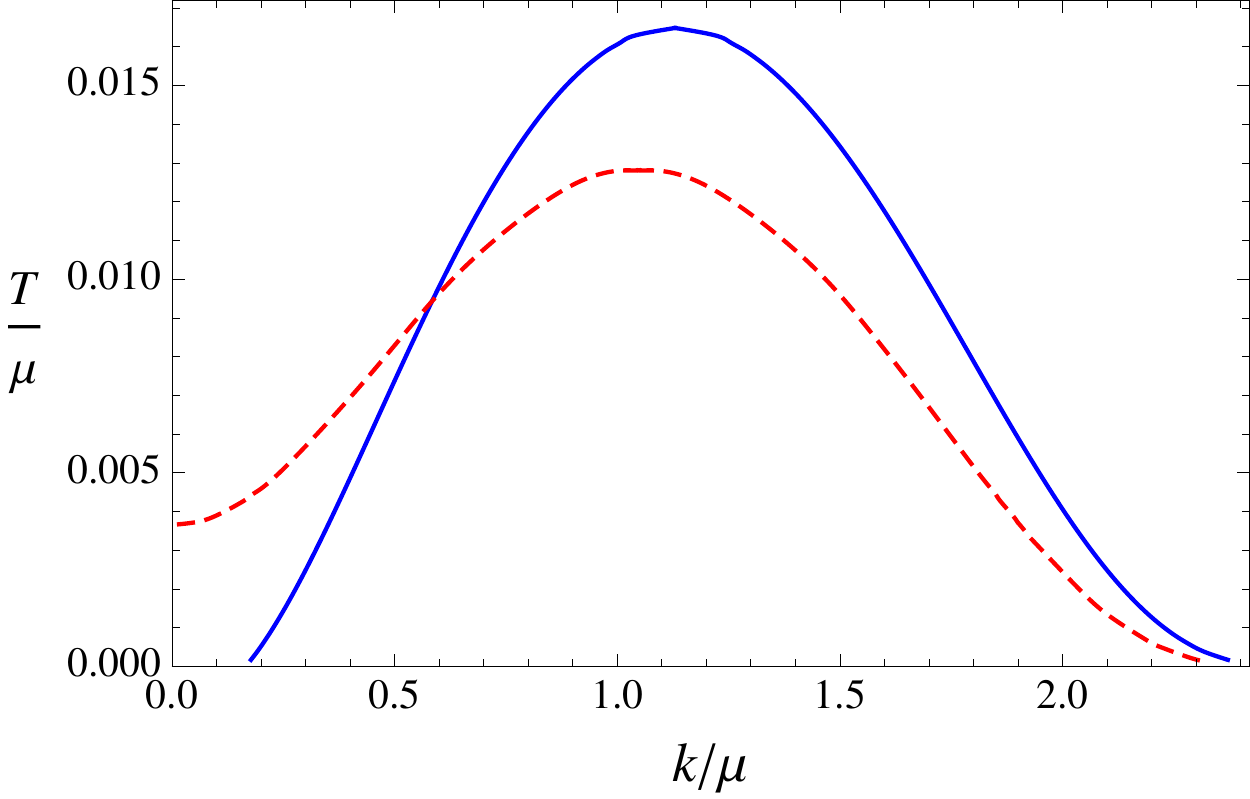}
\includegraphics[height=.153\textwidth]{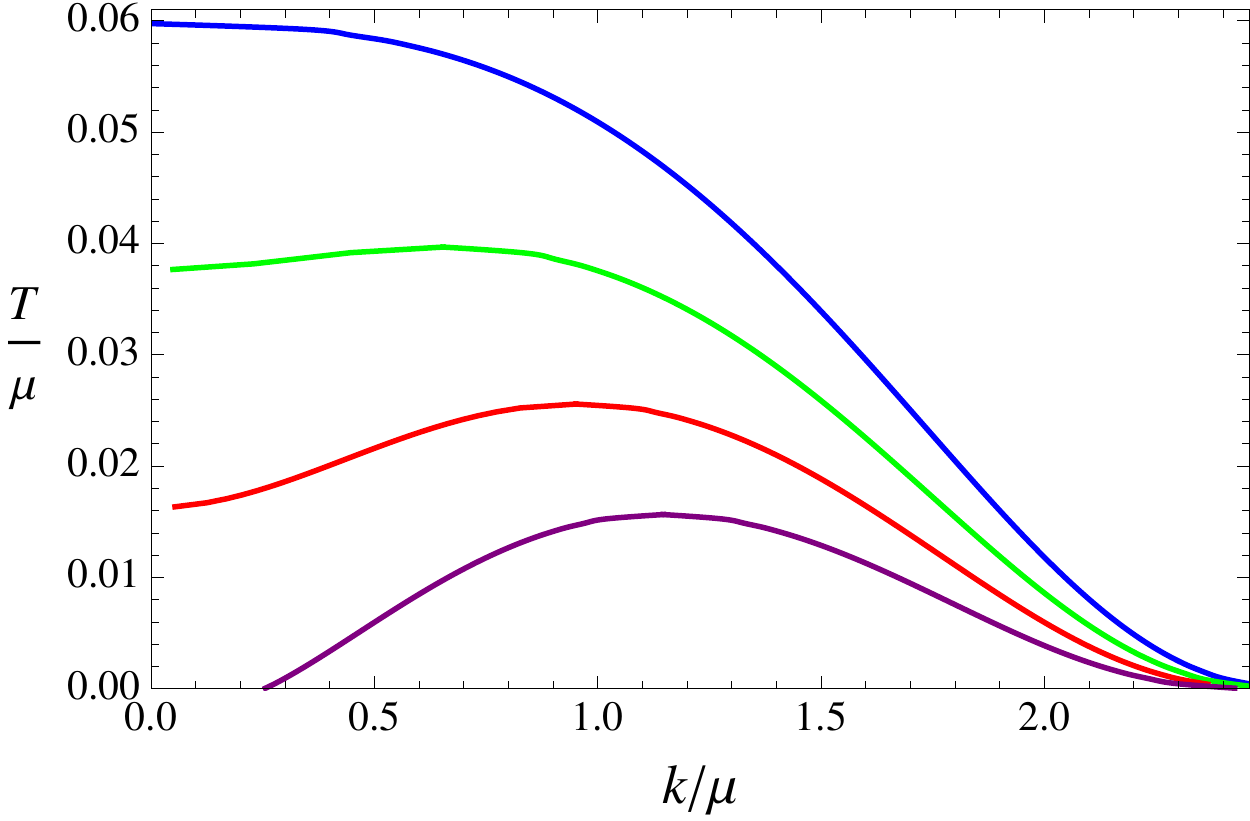}
\caption{Critical temperature as a function of wave number, for the onset of striped instabilities.
Left panel: the solid blue line describes the massless case $m_v^2=0$ with $|c|=2.34$, while the dashed red line the massive case $m_v^2=0.44$ with $|c|=2.46$. 
Right panel: dependence on the coupling $c$. From top to bottom $|c|=2.05, 2.15, 2.25, 2.35$.
In both figures the remaining parameters are chosen to be $m^2=-8, L=1/2, a=4, q_A=\mu=1$.}
\label{fig:tckcmv}
\end{center}
\end{figure}
For both scenarios we see the onset of a phase transition, as shown by the formation of a scalar condensate at $T_c$.

We show the dependence of the temperature on wave number in Fig.~1. 
In particular, the curves in the left panel exhibit clearly the bell curve behavior -- the fact that they are peaked 
at non-zero values of $k$ shows that the condensate is driven by the
momentum-dependent spatial modulations.
The right panel of Fig.~1 shows the dependence of $T_c$ on the strength $c$ of the coupling $Z_{AB}\sim c \, \chi$ between the two gauge fields. 
We would like to point out that as this coupling decreases, the effect of the spatial modulation also decreases -- one may still have a superconducting instability, but not striped.
Thus, in this model in order to ensure that the phase transition indeed occurs at finite values of $k$, the coupling must be non-zero and in fact sufficiently large. 
However, when $|c|$ becomes too large the instability once again disappears -- the BF bound can no longer be violated.

\section{Pair and Charge Density Waves}
In our model at low temperatures the scalar operator $\mathcal{O}_\chi$ dual to $\chi$ 
acquires a spatially modulated expectation value spontaneously, breaking the $U(1)$ symmetry.
Thus, the spatially modulated phase is always associated with a non-vanishing superconducting condensate. 
Moreover, the ``charge'' density $\rho_B$ associated with $B_\mu$ becomes spatially modulated,
and this, in conjunction with $\left<\mathcal{O}_\chi\right>$, induces a modulation in the charge density $\rho_A$ dual to $A_\mu$. 
While the second gauge field $B_\mu$ does not determine 
the type of order (PDW or SC+CDW) developed in the system, it can in principle be associated with spin degrees of freedom, with its modulated density $\rho_B$
describing SDW order.

As we have already mentioned, in a system with PDW order 
\begin{itemize}
\item
the average value of the superconducting order parameter $\langle O_\chi\rangle$ vanishes
\item
the charge density oscillations have half the period of those of the scalar condensate 
\end{itemize}
Thus, a PDW differs from a state with co-existing SC+CDW orders, in which the scalar condensate has a uniform component.
In our holographic model both of these features can be reproduced, along with the spontaneous -- and simultaneous -- breaking of the $U(1)$ symmetry and of translational invariance.
In particular, we find that when $q_B=0$ the scalar condensate and the charge density $\rho_A$ associated with the first vector field $A_\mu$ satisfy the conditions required for PDW order.
On the other hand, when $q_B \neq 0$ we find a state with SC + CDW order.

We have studied backreaction in our system numerically, focusing on the behavior of the scalar condensate $\left<\mathcal{O}_\chi\right>$ and of the two charge densities $\rho_A$ and $\rho_B$.
We work in the grand canonical ensemble by setting $\mu=1$ and
as an example, we choose the parameters in (\ref{coupling}) to be
$m^2=-8, m_v^2=0, L=1/2, c=-2.34, a=4, q_A=1$. We focus on the branch of solutions with $k=1$  and find a second order phase transition at $T_c= 0.01608$.
To gain intuition for our results, one can compare our numerics with a next-to-leading order perturbative analysis in $\varepsilon$, which in our case can be taken to be 
$\propto \sqrt{1-T/T_c}$ and  measures how close $T$ is to $T_c$,
\begin{eqnarray}
 \label{secondorder}
&&\delta \chi =\varepsilon\, w(r)\cos(k\,x)+\varepsilon^2[\chi^{(1)}(r)+ \chi^{(2)}(r)\cos(2k\,x)]\, , \nn \\ 
&&\delta B_t =\varepsilon\, b_t(r)\cos(k\,x)+\varepsilon^2[b_{t}^{(1)}(r)+ b_{t}^{(2)}(r)\cos(2k\,x)]\,,  \nn \\ 
&&\delta A_t = \varepsilon^2[a_{t}^{(1)}(r)+ a_{t}^{(2)}(r)\cos(2k\,x)]\,,
\end{eqnarray}
where we are singling out the perturbations of the scalar and vector fields for the sake of space.

We find that the order ${\cal O} (\varepsilon^2)$ components of $\delta \chi$ and $ \delta B_t $ are sourced by ${\cal O} (\varepsilon)$ terms proportional to $q_A \, q_B$, and therefore vanish when $q_B=0$.
In particular, note that this implies that the homogenous perturbations $\chi^{(1)}(r) $  and $ b_{t}^{(1)}(r)$ both vanish when $q_B=0$, causing the
scalar condensate $\langle O_\chi\rangle$ 
modulations (which to leading order are $\propto \cos (k\,x)$) 
to average out to zero. 
Note that by the same argument the oscillations of the charge density $\rho_B $ also average out to zero. Their period agrees with that of the scalar condensate, which is consistent with SDW order in a PDW.
On the other hand, since we are working at finite charge density with respect to 
$A_\mu$, the charge density $\rho_A$ always has a uniform component in our model.

Thus, the perturbative analysis suggests that the system behaves like a PDW when $q_B=0$ (no uniform component to $\langle O_\chi\rangle$), while when $q_B \neq 0$ it 
describes a SC+CDW state (the uniform contribution~$\propto \chi^{(1)}(r) $ is sourced).
This behavior is precisely confirmed by our numerics, as visible clearly in Fig.~2, which shows the oscillatory pattern of the scalar condensate $\langle O_\chi\rangle$ 
when $q_B=0$ (solid line) versus $q_B\neq 0$ (dashed line). In the former case the average value of the order parameter vanishes, but not in the latter.
\begin{figure}[ht!]
\begin{center}
\includegraphics[width=.32\textwidth]{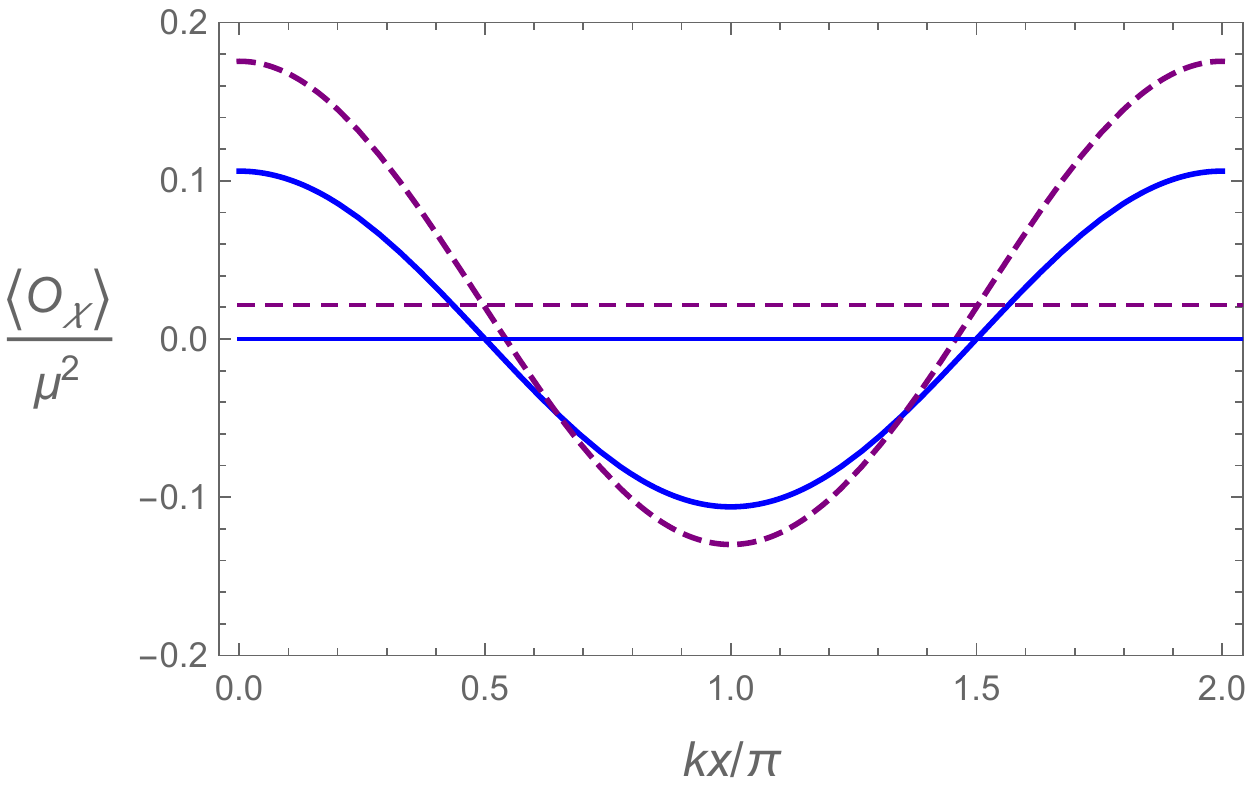}
\caption{
The scalar condensate for $T=0.01571$.
The solid blue curve corresponds to $q_B=0$, while the dashed purple line to $q_B=1/2$. 
The two horizontal lines denote the average values of the condensate in each case. Note that the average is zero only for $q_B=0$.}
\label{fig:O-x-qB}
\end{center}
\end{figure}

In order to have PDW order the period of the charge density must be one half of that of the scalar condensate. 
This is precisely what happens in our model when $q_B=0$, as shown in Fig.~3, 
where we clearly see that $\rho_A$ (dashed line) oscillates twice as fast
as the scalar condensate (solid line).
\begin{figure}[ht!]
\begin{center}
\includegraphics[width=.35\textwidth]{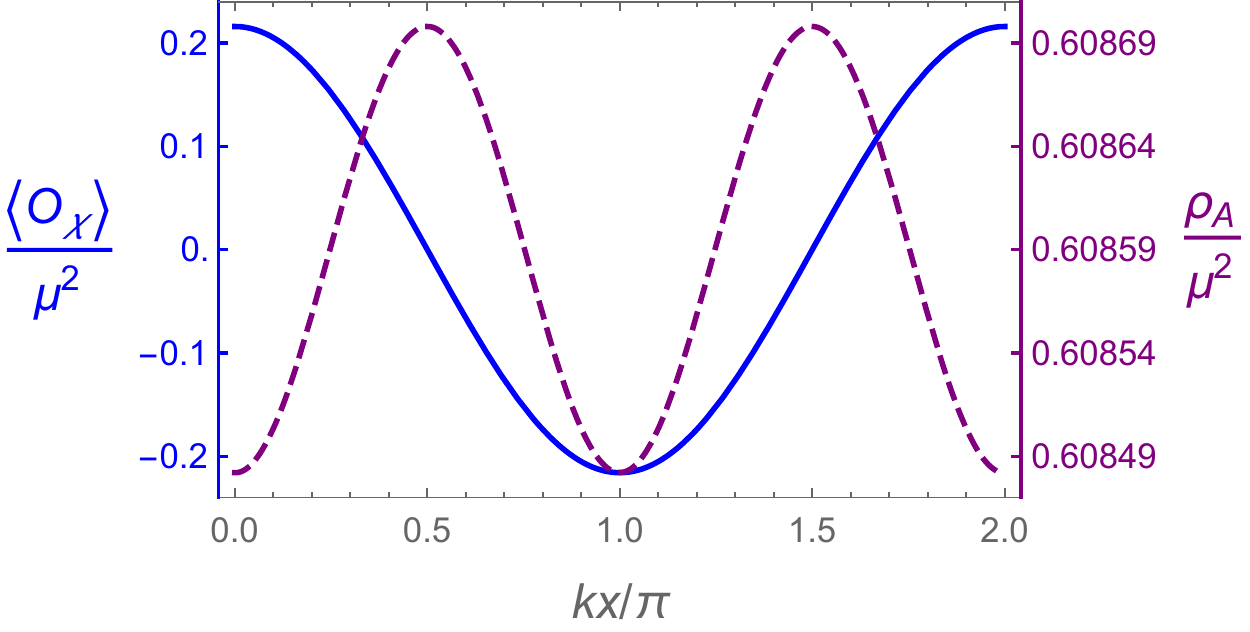}
\caption{The charge density $\rho_A$  (dashed purple line) associated with the $A_\mu$ gauge field, plotted against the scalar condensate (solid blue line) for $q_B=0$. 
The period associated with $\rho_A $ is one half of that of the scalar condensate.
We have chosen $T=0.01427$.}
\label{fig:O-x-qB}
\end{center}
\end{figure}

This result, which we have found numerically, can also be understood by inspecting the $\varepsilon^2 \cos (2k\,x)$ term in the perturbation $\delta A_t$, which is sourced 
by the product of the two ${\cal O}(\varepsilon)$ terms in $\delta\chi$ and $\delta B_t$.
Since the oscillation of $\rho_A$ is a next-to-leading order effect, which is sourced by the leading order oscillations of $\chi$ and $B_t$, this particular feature of the PDW order is in some sense~\emph{induced}.
We have also verified that for $q_B=0$ the frequency of the oscillations of the density $\rho_B$ is one half of that of $\rho_A$. We note that a similar doubling of frequencies was also seen in \cite{Donos:2016hsd} in the behavior of the 
magnetization densities.

On the other hand, when $q_B \neq 0$ the frequency of the oscillations of $\rho_A$ is the same as that of the condensate, which now has a uniform component.
Thus, what we have is a co-existing SC+CDW state, and not a PDW. 
This is shown clearly in Fig.~4, in which $\rho_A$ and $\langle O_\chi\rangle$ have the same period.
From the next-to-leading order perturbative analysis this is not quite clear. However, a $\cos (k\,x)$ mode is expected to appear at 
${\cal O} (\varepsilon^3)$ from the terms  $\omega(r) b_t^{(1)}(r) \cos (k\,x)  $ or $b_t(r) \chi^{(1)}(r) \cos (k\,x)  $ which are present when $q_B \neq 0$. 
Indeed, we have confirmed this in our numerics.

A more detailed analysis of this system will appear in~\cite{LongerPaper}, where we will include the behavior of the background geometry and the thermodynamics.
While in this paper the matter content was chosen for its simplicity, more complicated models with additional fields can in principle be constructed, 
in which the superconducting state is associated with the condensation of a~\emph{complex} scalar.
Moreover, at very low temperatures the physics encoded in this model may be richer than what our preliminary analysis has shown.
A very interesting question is that of the nature of the ground state once striped superconducting order develops.
Finding the fully backreacted geometry at zero temperature remains a challenge.
\begin{figure}[ht!]
\begin{center}
\includegraphics[width=.35\textwidth]{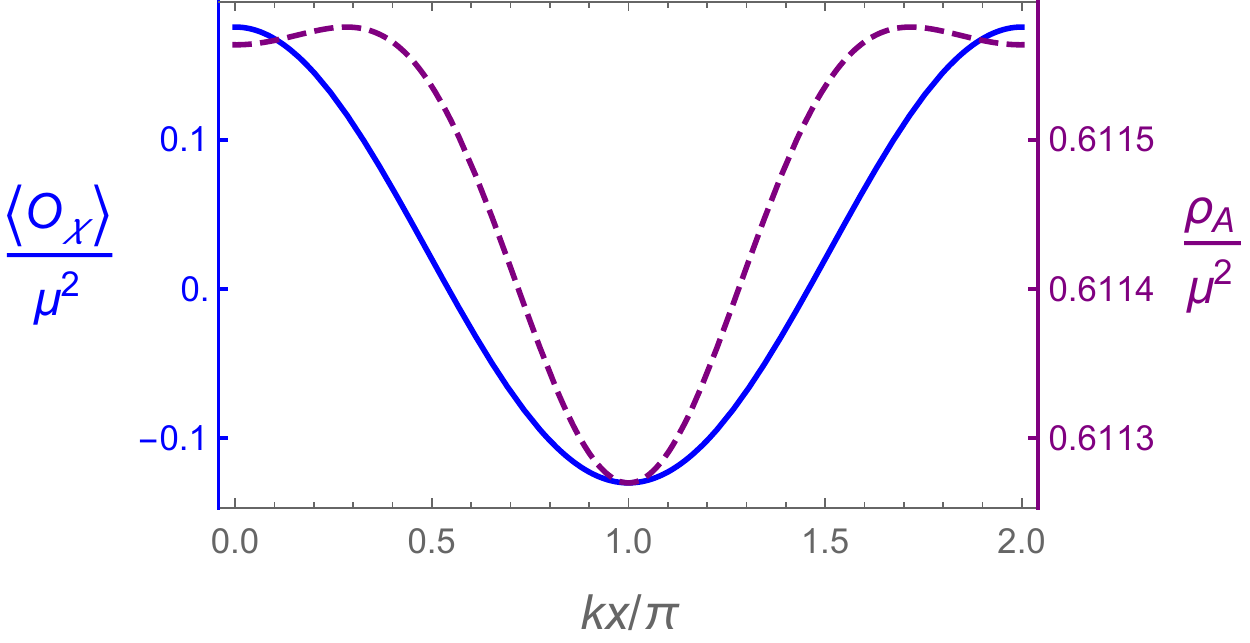}
\caption{The scalar condensate (solid blue line) and charge density $\rho_A$ (dashed purple line) for $q_B=1/2$ at $T=0.01571$.
The two share the same period.}
\label{fig:O-x-qB}
\end{center}
\end{figure}
%



\section{Acknowledgments}
We would like to thank R.G. Cai, E. Kiritsis and S. Kachru for useful discussions.
J.R. is partially supported by the American-Israeli Bi-
National Science Foundation, the Israel Science Foundation
Center of Excellence and the I-Core Program of the
Planning and Budgeting Committee and The Israel Science
Foundation ``The Quantum Universe.''
The work of S.C. is supported in part by the National Science Foundation grant PHY-1620169.\\

\end{document}